\documentclass[aps,floatfix,twocolumn,a4paper,showpacs, nofootinbib,superscriptaddress,10pt]{revtex4}
\usepackage{graphicx,float}\usepackage{graphicx,float}
\usepackage[all]{xy}
\usepackage{amsmath,upgreek}
\usepackage{amssymb}
\usepackage{color}
\usepackage{epsfig,bm}		
\usepackage{graphicx,epstopdf}
\usepackage{subfigure}
\usepackage{pdfpages,slashed}
\usepackage[colorlinks,hyperindex]{hyperref}

\definecolor{green1}{RGB}{0,128,0}
\hypersetup{backref=true,pagebackref=true}
\hypersetup{%
  colorlinks = true,
  linkcolor  = blue,
  citecolor = cyan,
}
\usepackage{bookmark,textgreek}
\usepackage{hyperref,color,xcolor}
\hypersetup{hidelinks,hyperindex=true,colorlinks=true,breaklinks=true,urlcolor= blue}
\hypersetup{%
  colorlinks = true,
  linkcolor  = blue
}\usepackage{amssymb}
\newsavebox{\foobox}

\usepackage{graphicx,float,tikz}
\usepackage[all]{xy}
\newcommand\ringring[1]{%
  {
   \mathop{\kern0pt #1}\limits^{
     \vbox to-1.85ex{
       \kern-2ex 
       \hbox to 0pt{\hss\normalfont\kern.1em \r{}\kern-.45em \r{}\hss}%
       \vss 
     }
   }
  }
}\newcommand\orcidroldao{{\href{https://orcid.org/0000-0003-3978-532X}{\orcidicon}}}
\newcommand{\orcidicon}{%
	\begin{tikzpicture}
	\draw[lime, fill=lime] (0,0)
		circle [radius=0.16]
		node[white] {{\fontfamily{qag}\selectfont \tiny ID}};
	\draw[white, fill=white] (-0.0625,0.095)
		circle [radius=0.007];
	\end{tikzpicture}	\hspace{-2mm}
}
\newcommand{\bpartial}{\mathop{\partial\kern -4pt\raisebox{.8pt}{$|$}}}

\newcommand{\bes}{\begin{subequations}}
\newcommand{\ees}{\end{subequations}}
\def\beq{\begin{eqnarray}}
 \newcommand{\clt}{\textcolor{black}}
  
\def\eeq{\end{eqnarray}}
\def\be{\begin{equation}}
\def\ee{\end{equation}}

\begin{document}

\title{Configurational entropy of heavy-quark QCD exotica}
\author{G. Karapetyan}
\email{gayane.karapetyan@ufabc.edu.br}
\affiliation{Federal University of ABC, Center of Natural Sciences, Santo Andr\'e, 09580-210, Brazil}
\affiliation{Federal University of ABC, Center of Mathematics, Santo Andr\'e, 09580-210, Brazil}
\author{R. da Rocha\orcidroldao\!\!}
\email{roldao.rocha@ufabc.edu.br}
\affiliation{Federal University of ABC, Center of Mathematics, Santo Andr\'e, 09580-210, Brazil}

\begin{abstract}
The mass spectroscopy of exotic meson states is scrutinized in the AdS/QCD paradigm. The differential configurational entropy is then used to study, derive, and analyze the mass spectrum of excited exotic vector meson resonances, whose configurational stability is also addressed. For it the hadronic molecule, the hadrocharmonium, and the hybrid description of heavy-quark exotic mesonic states
are employed to more precisely match experimental data in the Particle Data Group and also to predict the next generations of heavy-quark QCD exotica in the $Z_c$, $Y$, $\pi_1$, and $Z_b$ exotic meson families.

 \end{abstract}
\maketitle

\section{Introduction}

Information entropy, innovatively proposed by Shannon, can resolve the uncertainty within a physical system. Based on communication, it can be implemented by an array of messages that are sent through noisy channels and subsequently reconstructed with the lower error probability distribution \cite{Shannon:1948zz}. 
Information entropy is inherent to probability distributions, being the precise rate limit at which one can compress data promoted by an independent random variable. It also represents the maximum rate of message transmission along noisy channels. 
The configurational entropy (CE) implements the information entropy in discrete physical systems, representing the way how information can be cyphered into wave resonance modes in the space of momenta, in such a way that the entire physical system can be described nonetheless \cite{Gleiser:2011di,Gleiser:2012tu}. The differential configurational entropy (DCE) takes place in the study of continuous systems \cite{Gleiser:2018kbq}, being pivotal to describe particles and resonances in quantum chromodynamics (QCD) \cite{Bernardini:2016hvx,Karapetyan:2018oye,Bernardini:2018uuy}.

The anti-de Sitter/QCD (AdS/QCD) duality implements the non-perturbative confinement mechanism of QCD, where quarks are confined within hadronic states. Weakly-coupled gravity in AdS${}_5$ bulk consists of a dual description
of non-perturbative QCD encoding strong interactions \cite{Karch:2006pv,Branz:2010ub,Colangelo:2008us,Brodsky:2014yha,EKSS2005}. This description comes from the  conjecture by Maldacena, asserting that the large-$N_c$ regime of the $\mathcal{N} = 4$ SU($N$) super  
Yang-Mills theory in a 4D spacetime corresponds to the type IIB string theory
in AdS${}_5 \times S^5$ compactification \cite{Maldacena:1997re}. It accommodates a consistent strongly-coupled QCD theory, living on the AdS${}_5$ boundary, in the low energy regime. The AdS${}_5$ bulk accommodates weakly-coupled gravity as the QCD dual sector. The codimension represents the energy scale in QCD. In this limit, bottom-up AdS/QCD approaches can emulate relevant outcomes of QCD by appropriate choices of warped metrics endowing the AdS${}_5$ bulk and suitable dilaton backgrounds.

AdS/QCD has been thoroughly explored under the DCE tools, mainly scrutinizing hadronic properties, also proposing new aspects of hadronic resonances as well as corroborating experimental data. Several light- and heavy-flavor meson families, including isovector, axial vector, vector, pseudoscalar, scalar, and tensor ones; baryonic resonances, glueball fields, and reggeons have been investigated at zero and finite temperatures, employing the DCE underlying AdS/QCD \cite{Ferreira:2019inu,daRocha:2021ntm,daRocha:2021imz,Ferreira:2019nkz,Barbosa-Cendejas:2018mng,Braga:2018fyc,Colangelo:2018mrt,Ferreira:2020iry,MarinhoRodrigues:2020yzh,Bernardini:2016qit,Braga:2017fsb,Karapetyan:2018yhm,Ma:2018wtw}. 
Also, new mesonic states have been proposed and matched to detected resonances reported in the Particle Data Group (PDG) \cite{pdg}, with the aid of the DCE in AdS/QCD. The DCE has been also used to probe other several aspects of QFT  \cite{Bazeia:2018uyg,Bazeia:2021stz,Correa:2016pgr,Cruz:2019kwh,Lee:2019tod,Alves:2014ksa,Alves:2017ljt,Gleiser:2018jpd,Braga:2016wzx,Fernandes-Silva:2019fez}.

Seminal developments in AdS/QCD support universal properties that comprise mesonic states from light-flavor ones to heavy bottomonia and charmonia \cite{Badalian:2016ttl,dePaula:2009za,Braga:2017oqw,Bernardini:2003ht}, also including open charm and open beauty mesons states. 
Some models in AdS/QCD report quadratic dilatons  \cite{MartinContreras:2020cyg,Song}, wherein Regge trajectories can precisely portray the (squared) mass spectrum of light-flavor mesonic states as linear functions of the excitation number of $S$-wave mesonic states \cite{Rinaldi:2020ybv,Rinaldi:2020ssz}. For the case of mesons composed of heavy quarks, the linearity of Regge trajectories does not hold anymore, although the heavy meson spectrum can be still described with precision \cite{Afonin:2014nya,jk}.
Also, heavy quarks can compose the so-called heavy-quark exotica, like tetraquarks, quarkonium-like exotic mesons, hadrocharmonium, hadronic molecules, and hybrids \cite{Guo,Lebed1}, which will be scrutinized here in the context of the DCE. 

Among the plethora of already detected particles and resonances, exotic mesons remain a partially disassembled puzzle in QCD when one attempts to model and interpret their fundamental properties. Exotic mesonic states do not obey the standard $q\bar q$ structure, where $q$ denotes a quark \cite{Lebed1,Jaffe1}. The standard constituent quark model reports detected mesons as bound $q\bar q$ states clustered 
into SU($N$) (flavor) multiplets. Nevertheless, the gluonic coupling in QCD indicates that mesons can be also composed of $q\bar q$ states bounded to a gluon. These states are called hybrids. Besides, multiquark color
singlet states of type $qq \bar q\bar q$ encompass tetraquarks and also hadronic molecules that are two bounded mesons,  
with experimental data reported in PDG 2020 \cite{pdg}. They are experimental confirmation of heavy-quark states afar the standard quark model \cite{Guo,Liu}. Some of these states, 
do not fit into the conventional $c\bar c$
 spectrum (where $c$ denotes the charm quark), and are named charmonium-like exotic meson states. What discerns multiquark states containing a $c\bar c$ pair from
conventional charmonia is the existence of multiplets that include states with non-vanishing charge (as the $c\bar c u\bar d$), non-vanishing strangeness (as the $c\bar c d\bar s$), or even both (as the $c\bar c u\bar s$) \cite{Godfrey:2008nc}. An analogous definition holds for bottomonium-like states. There are plenty of other exotic composites formed by heavy-flavor quark states, as well as their appropriate fusion \cite{Abreu:2017nuc}.

Hadronic molecules are also fundamental exotic states consisting of a pair of meson states that are bound by the strong force, also interacting via a residual weak QCD colorless force. They can be also realized either as two heavy quarkonia or one heavy quarkonium bounded to a light-flavor mesonic state. 
Currently, several models shed light on the behavior of such exotics. Nevertheless, even such theories can neither completely cover the existing experimental data nor explain still several properties of exotic mesons.

The DCE of vector heavy-quark exotic mesons will be evaluated in this work to probe the mass spectra of the next generation of resonances in each of the heavy-quark exotic meson $Z_c$, $Y$, $\pi_1$, and $Z_b$ families. To accomplish it, the DCE will be calculated for the excited $S$-wave resonances as well as a function of the exotic mesonic states experimental mass spectrum. The procedure to be used is based on the interpolation of DCE Regge-type trajectories that are engendered using the experimental mass spectrum of each one of the $Z_c$, $Y$, $\pi_1$, and $Z_b$ exotic meson families. They comprise candidates in the next generations proposed to be detected in experiments, as well as candidates that match up to states in PDG \cite{pdg}. 
This work is organized as follows: Sec. \ref{sec1} is devoted to introducing heavy-quark exotic meson states and how to derive their mass spectrum of each $Z_c$, $Y$, $\pi_1$, and $Z_b$ exotic meson families, using an AdS/QCD soft wall-based dilaton that takes into account the massive heavy quarks that compose exotic mesons. 
In Sec. \ref{sec2} the DCE paradigm is discussed and evaluated for the $Z_c$, $Y$, $\pi_1$, and $Z_b$ exotic meson families, concerning $S$-wave excitations and the experimental mass spectrum of each meson family, whose interpolation yields DCE Regge-type trajectories.
The procedure of extrapolating the obtained  trajectories generates the mass spectrum of heavier exotic meson excitations, in every one of the exotic meson families. The DCE also infers the configurational stability analysis of these exotic mesonic excitations.  Sec. \ref{iv} concludes and discusses the most relevant results, proposing important perspectives and extensions as well.

\section{Heavy-quark QCD exotica in AdS/QCD}
\label{sec1}

Meson states having quantum numbers that are forbidden by the standard $q\bar{q}$ paradigm are denominated exotic mesons. 
They comprise, in particular, charmonium-like charged resonances in the $Z_c$ exotic meson family with $I^G(J^{PC})=1^+(1^{+-})$. Resonances in the $Z_c$ family present more valence quarks, as the $Z_c(4430)$, in contrast to the neutral charmonium $c\bar c$. One of the first discovered exotic mesons was the $Z_c(3900)^+$, formed by a $c\bar c u\bar d$ state \cite{Ablikim}. Quarkonium-like states are labeled by $Z$ and in general, can have neutral isospin partners. For instance, $Z_c(3900)^+$ has a neutral counterpart $Z_c(3900)^0$ state \cite{Ablikim2}. Bottomonium-like states are labeled by $Z_b$, whereas charmonium-like ones carry the notation $Z_c$. The molecular portray of multiquark states can be also approached. For example, a tetraquark state can be described as a hadronic molecule composed of four valence quarks into two mesons. There are several distinct ways to constitute a color singlet of either $qq\bar{q}\bar{q}$ or composition $Qq\bar{Q}\bar{q}$, where $Q$ denotes hereon either a charm or a bottom heavy quark \cite{Lebed1}. 
Within a tetraquark the color-singlet pairs can be delocalized, as the example of the $X(3872)$ exotic state described as a hadronic molecule. 

Exotic mesons also encode hadroquarkonium states, which are composed of a heavy-quark pair $Q\bar Q$ and a light-flavor $q\bar q$ vector meson cloud \cite{Brodsky}.
Each quark composite engenders a color singlet, glued through weak color van der Waals forces. From the phenomenological point of view, the standard charmonium decay, mainly consisting of the $J/\psi$, $\psi(2S)$, and $\chi_c$ resonances, takes part of the dissociation of the charmonium core from the light-flavor cloud \cite{Campanella:2018xev}. 
An important example is the $Z_c(4430)$ decay to $\psi(2S)$, in comparison to the $\psi(1S)=J/\psi$ state, that can be realized as a yield of a 
 $c\bar c$ core in $Z_c(4430)$. 
The sequence of states $Y(4008)$, $Y(4230)$, $Y(4260)$, $Y(4360)$, and $Y(4660)$, whose three last members will be studied in the DCE and AdS/QCD contexts in Sec. \ref{sec2}, represent heavy-quark exotic states, splitting into standard charmonium and light-flavor mesons.
Hadrocharmonium involves a composite of compact charmonium and light-flavor quarks, the $c\bar c$ playing the role of the system core. The composite has $I^G(J^{PC}) = 0^+(1^{--})$ quantum numbers.  
The simplest hadrocharmonium comprises a meson pair that are extended over, emulating in QCD a diatomic molecule-like picture \cite{Lebed1}.

Going beyond the non-$q\bar q$ elements in the meson family, the so-called exotic hybrids, that also will be used in Sec. \ref{sec2}, include else valence\footnote{It is worth mentioning that, as emphasized in Ref. \cite{Lebed1}, valence gluons do influence the $J^{PC}$ quantum numbers in hadrons, contrary to other created and annihilated gluons, without even taking into account the virtual quark-antiquark pairs. } gluons as degrees of freedom, additionally to quarks constituents \cite{Meyer}.  
Hybrids are also realized as pairs of quarks and antiquarks joined by gluonic flux tubes. This particular configuration permits one to regard other quantum number values that are forbidden in the $q\bar{q}$ paradigm, as the case of exotic meson states
 $\pi_1$, with $I^G(J^{PC})=1^-(1^{-+})$ and bottomonium-like states $Z_b$, with $I^G(J^{PC})=1^+(1^{+-})$. 
 
Heavy-quark QCD exotica can be implemented in the context of AdS/QCD, by adapting the AdS/QCD soft-wall \cite{Karch:2006pv} in such a way that massive quarks can be taken into account \cite{MartinContreras:2020cyg}. One regards a vector field $E_M = (E_\mu,E_z)$ in the AdS${}_5$ bulk space under the rules of the action
\begin{equation}
\mathfrak{I} = -k_5^2 \int  e^{-\Upphi (z)}\mathfrak{L} \,\sqrt{-|g|}\,d^5x,
\label{vectorfieldaction}
\end{equation} where $k_5^{2}$ is the effective constant coupling and 
\beq\label{lalag}
\mathfrak{L}=\frac12 g_{PQ}F^{MP} F^{NQ}\eeq
 denotes the Lagrangian, for the Yang-Mills field strength $F_{MN} = \partial_{[M}E_{N]}$.
 The metric
\begin{equation}\label{space1}
 g_{MN}dx^M dx^N= e^{2A(z)}\left(dz^2+\upsigma_{\mu\nu}dx^\mu dx^\nu\right)\,
\end{equation}
endows the AdS${}_5$ space, where the warp factor reads ${A}(z) = -\ln\left(\frac{z}{R}\right)$ and $\upsigma_{\mu\nu}$ is the metric with signature ($-,+,+,+$) that equips the AdS${}_5$ boundary. Analogously to standard mesons in AdS/QCD, exotic mesonic states can be also characterized via the conformal dimension $\Updelta$, that is constrained to both the bulk vector field mass, $\mathtt{M}$, and the spin by 
 \cite{Witten:1998qj}
\begin{equation}\label{mesao}
 \mathtt{M}_5^2 R^2=(\Updelta+p-4)(\Updelta-p). \end{equation}
 Ref. \cite{MartinContreras:2020cyg} proposed 
 the dilaton
\beq
\Upphi(z) = (\kappa z)^{2-\beta},\label{devia}
\eeq where the energy scale $\kappa$ is the scale energy and defines the angle of the associated Regge trajectory and $\beta$ encrypts the information about heavy massive quarks. It extends the standard soft wall quadratic dilation, that can be recovered in the $\beta\to0$ limit, that takes into account massless quarks and is appropriate to describe light-flavor mesons \cite{Karch:2006pv,Brodsky:2014yha}.

The Euler-Lagrange equations applied to the action (\ref{vectorfieldaction}) yield an EOM ruling heavy-quark vector exotic mesons,  
\begin{equation}\label{eom1}
\left[\partial_z^2-{\rm B}'(z)\partial_z-q^2\right]E_\mu(z,q)=0,   
\end{equation}
\noindent where ${\rm B}(z)+A(z)=\Upphi(z)$ and ${\rm B}'(z)=d{\rm B}(z)/dz$. With the field decomposition 
$
E_\mu(z,q)=E_\mu(q)\,\upchi(z,q)$, Eq. (\ref{eom1}) can be further reduced by the Bogoliubov splitting $\upchi(z,q)=e^{\Upphi(z)/2}\xi(z,q)$  \cite{MartinContreras:2020cyg} 
\begin{equation}\label{schr}
\left[-\partial_z^2+\mathbb{U}(z)\right]\,\xi(z,q)=-q^2\xi(z,q),   
\end{equation}
with potential 
\beq\label{ppp}
\!\!\!\!\!\!\!\!\!\!\!\!\mathbb{U}(z,\kappa,\beta)&=&\frac{3}{4 z^2}+\left(\frac{3\beta}2-\frac{\beta ^2}{2}  -\kappa ^2\right)\kappa^2 (\kappa z)^{-\beta }\nonumber\\
&&+\frac{\kappa^2}{4}\left(2-\beta\right)^2(\kappa z)^{2-2 \beta }\nonumber\\
&&+\frac{\kappa }{z}\left(1-\frac{\beta}{2}\right)(\kappa z)^{1-\beta }+\frac{\mathtt{M}_5^2(\Updelta)\,R^2}{z^2}.
\eeq
\noindent  
The Schr\"odinger-like equation (\ref{schr}) is then responsible for obtaining the mass spectrum, $m_n^2 = -q^2$, of vector heavy-quark exotic meson states $\xi_n(z,q)$. \clt{For it, determination of the parameters $\kappa, \beta$, and $\Updelta$ match the experimental mass spectrum of the detected elements in these families \cite{MartinContreras:2020cyg,Liu}. Each $Z_c$, $Y$, $\pi_1$, and $Z_b$ exotic meson family can be uniquely characterized by the conformal dimension $\Updelta$,  determined by the dual bulk operators, and constrained to both the AdS bulk vector field mass and the AdS radius \cite{Witten:1998qj}. For each one of the $Z_c$, $Y$, $\pi_1$, and $Z_b$ exotic meson families, $\Delta$ is fixed and determined by the dual bulk operators.} 
To obtain the mass spectra of these families that more precisely match data in PDG 2020 \cite{pdg}, $Z_c$ exotic meson states will be described by hadronic molecules,  
the $Y$ exotic meson family is better described as hadrocharmonium states, 
and, the $\pi_1$ and $Z_b$ exotic families are precisely reported when the hybrid heavy-quark prescription is taken into account, among other possible choices \cite{MartinContreras:2020cyg}. 
Complementary to \eqref{eom1}, there is a massive EOM governing the $z$-component, 
\begin{equation}\label{boxx}
\left(\Box+\mathtt{M}_5^2e^{2A}\right)E_z-\partial_z\left(\slashed{\partial}\cdot E\right)=0, 
\end{equation}
where the notation $\slashed{\partial}\cdot E=\partial_\mu\,E^\mu$ is used.
The gauge condition $E_z=0$ hence yields the Lorentz gauge 
$\slashed{\partial}\cdot E=0$, with transverse boundary fields. 
As the conformal dimension encodes the number of constituent quarks, it does not differentiate among hadroquarkonia, hadronic molecules or hybrid exotica \cite{MartinContreras:2020cyg}. However, one can split  
heavy-quark meson exotic states into these families when the mass of each state is taken to be a weighted sum of quarks $(q_k)$, and mesons, 
\beq\label{sumi}
\bar{m}=\sum_{k=1}^N\left(p_k^{q_k}\bar{m}_{q_k}+p_k^{\scalebox{.67}{\textsc{meson}}} m_{{\scalebox{.67}{\textsc{meson}}}}\right), 
\eeq
with $\sum_{k=1}^N\left(p_k^{q_k}+p_k^{{\scalebox{.67}{\textsc{meson}}}}\right)=1$, where  $p_k^{q_k} \, [p_k^{{\scalebox{.67}{\textsc{meson}}}}]$ denote the respective weights for the quarks [mesons].

Ref. \cite{41} suggests that $Z_c$ exotic meson states can be described by the hadronic molecule ({\scalebox{0.9}{\textsc{HC}}}) picture, containing at least one pair of $c\bar{c}$ in the molecule core. The threshold mass for the holographic $Z_c$ exotic mesons is 
\beq\label{sumi1}
\bar{m}_{_{\scalebox{.67}{\textsc{HM}}}} = 0.283 m_{J/\psi} + 0.717 m_\rho,\eeq as suggested by Ref. \cite{MartinContreras:2020cyg}. 
The resulting data are respectively encoded in Tables \ref{scalarmasses1} -- \ref{scalarmasses4}.
\begin{table}[h]
\begin{center}
\begin{tabular}{||c|c||c|c||}
\hline\hline
$n$ & State & $M_{\scalebox{.65}{\textsc{Experimental}}}$ (MeV) & $M_{\scalebox{.65}{\textsc{Theory}}}$ (MeV) \\
    \hline\hline
\hline
1 &\;$Z_c(3900)$ & $3888.4\pm2.5$ & $3816.3$ \\ \hline
2 &\;$Z_c(4200)$ & $4196.0^{+31}_{-29}$ & $4213.9$ \\ \hline
3& \;$Z_c(4430)$ & $4478.0^{+15}_{-18}$ & $4551.4$ \\\hline
\hline\hline
\end{tabular}
\caption{Mass spectrum of $Z_c$ exotic meson resonances, in the hadronic molecule prescription. Regarding the masses in the fourth column derived by solving Eq. \eqref{schr}, the parameters $\Delta=6$, $\beta= 0.539$, and $\kappa = 2.151$ GeV are adopted. } \label{scalarmasses1}
\end{center}
\end{table}
\noindent


\begin{table}[h]
\begin{center}\begin{tabular}{||c|c||c|c||}
\hline\hline
$n$ & State & $M_{\scalebox{.65}{\textsc{Experimental}}}$ (MeV) & $M_{\scalebox{.65}{\textsc{Theory}}}$ (MeV) \\
    \hline\hline
\hline
1 &\;$Y(4260)$ & $4209.1\pm6.8$ & $4228.3$ \\ \hline
2 &\;$Y(4360)$ & $4368.0\pm13.0$ & $4577.3$ \\ \hline
3 & \;$Y(4660)$ & $4633.0\pm7.0$ & $4871.8$ \\\hline
\hline\hline
\end{tabular}
\caption{Mass spectrum of $Y$ exotic meson resonances, in the hadrocharmonium prescription. Regarding the masses in the fourth column derived by solving Eq. \eqref{schr}, the parameters $\Delta=6$, $\beta= 0.604$, and $\kappa = 2.523$ GeV are adopted. } \label{scalarmasses2}
\end{center}
\end{table}
The $Y(4260)$, $Y(4360)$, $Y(4660)$ masses are not
compatible with any of the charmonium states and can be well described using hadrocharmonium ({\scalebox{0.9}{\textsc{HC}}}) \cite{Li:2013ssa}. From the AdS/QCD point of view, the operator creating these states has conformal dimension $\Updelta = 6$, yielding a bulk mass such that  
$\mathtt{M}_5^2 R^2 = 15$. A charmonium core characterized by its mass and the light-flavor quark cloud as well, consisting of a pair of $u$ and $d$ quarks, will be regarded and the holographic threshold suggesting the quark constituent mass reads \cite{MartinContreras:2020cyg}
\beq\label{sumi2}
\bar{m}_{\scalebox{.67}{\textsc{HC}}}= \frac12 m_{J/\psi} + \frac14(\bar{m}_u + \bar{m}_d).\eeq

In hybrid exotic mesons ({\scalebox{1}{\textsc{hybm}}}), gluon(s) effectively composes the mesonic structure, instead of being just an exchange particle whose interaction bound the quarks together. \clt{To engender the holographic description, one needs to
define the hadronic operators creating hybrid mesons, from the AdS/QFT dictionary, that uses 2-point functions at the conformal boundary. These objects are defined in terms of operators that are composites of quarks
and gluons. Hybrid exotic mesons are $q\bar q$ states bounded to a gluon.} The quark-gluon composite operator read $Q \upgamma_\mu \bar{Q}\mathbb{G}^{\mu\nu}$, where $\mathbb{G}^{\mu\nu}$ is a ground state gluonic field. It implies that $\Updelta = 5$ [$\Updelta = 7$] for one [two] composite gluons, implying that $\mathtt{M}_5^2 R^2 = 8 \;[24]$, where the gluon composite has mass $M_g \approx 0.7$ GeV \cite{MartinContreras:2020cyg}. Hence, denoting by $m_g$ tthe gluon(s) mass, one can write 
\beq\label{hybm}
\bar{m}_{\scalebox{.67}{\textsc{hybm}}} = p_Q m_Q + p_{\bar{Q}} m_{\bar{Q}} + p_g m_g.\eeq 
For describing the $\pi_1$ exotic meson states, $p_Q=0.497=p_{\bar{Q}}$ and $p_g = 0.006$, whereas to represent the $Z_b$ exotic mesons one takes $p_Q=0.49=p_{\bar{Q}}$ and $p_g = 0.01$ \cite{MartinContreras:2020cyg}.

\begin{table}[h]
\begin{center}
\begin{tabular}{||c|c||c|c||}
\hline\hline
$n$ & State & $M_{\scalebox{.65}{\textsc{Experimental}}}$ (MeV) & $M_{\scalebox{.65}{\textsc{Theory}}}$ (MeV) \\
    \hline\hline
\hline
1 &\;$\pi_1(1400)$ & $1354.0\pm25.0$ & $1351.7$ \\ \hline
2 &\;$\pi_1(1600)$ & $1660.0^{+15}_{-11}$ & $1646.6$ \\ \hline
2 &\;$\pi_1(2015)$ & $2014.0\pm20.0$ & $1901.7$ \\ \hline
\hline\hline
\end{tabular}
\caption{
Mass spectrum of $\pi_1$ exotic meson resonances, in the hybrid meson prescription. Regarding the masses in the fourth column derived by solving Eq. \eqref{schr}, $\Delta=5$, $\beta= 0.036$ and $\kappa = 0.488$ GeV. } \label{scalarmasses3}
\end{center}
\end{table}
\noindent

\begin{figure}[h]
	\centering
	\includegraphics[width=6.8cm]{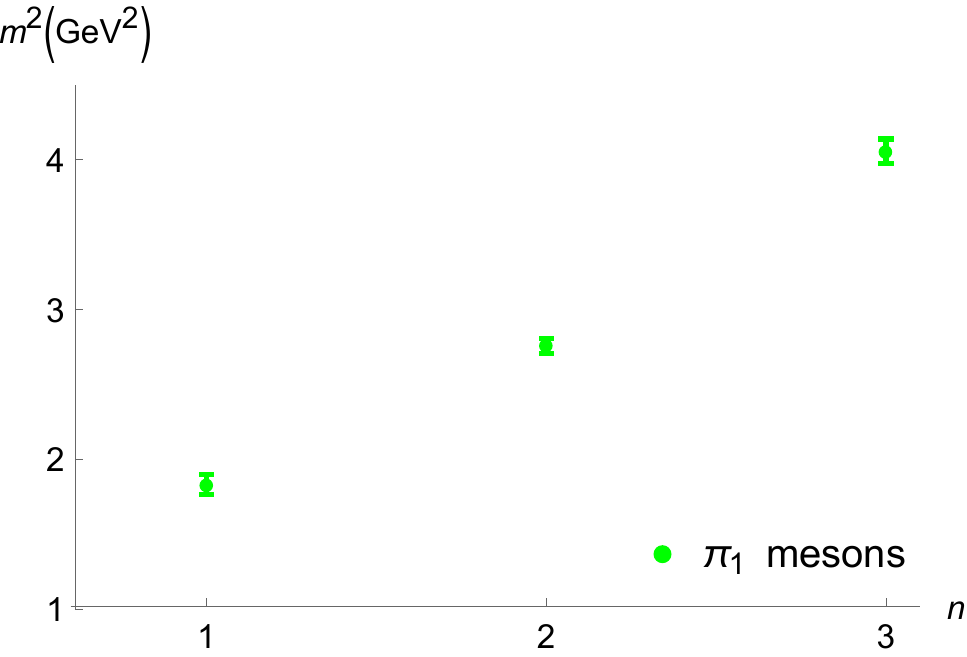}
	\caption{Experimental mass spectrum of $\pi_1$ exotic mesonic states as a function of radial excitations, for $n=1,\ldots,3$ (respectively corresponding to the $\pi_1(1S) =\pi_1(1400)$, $\pi_1(2S) = \pi_1(1600)$ and $\pi_1(3S) = \pi_1(2015)$ states.}
	\label{tudonmn}
\end{figure}

\begin{figure}[h]
	\centering
	\includegraphics[width=6.8cm]{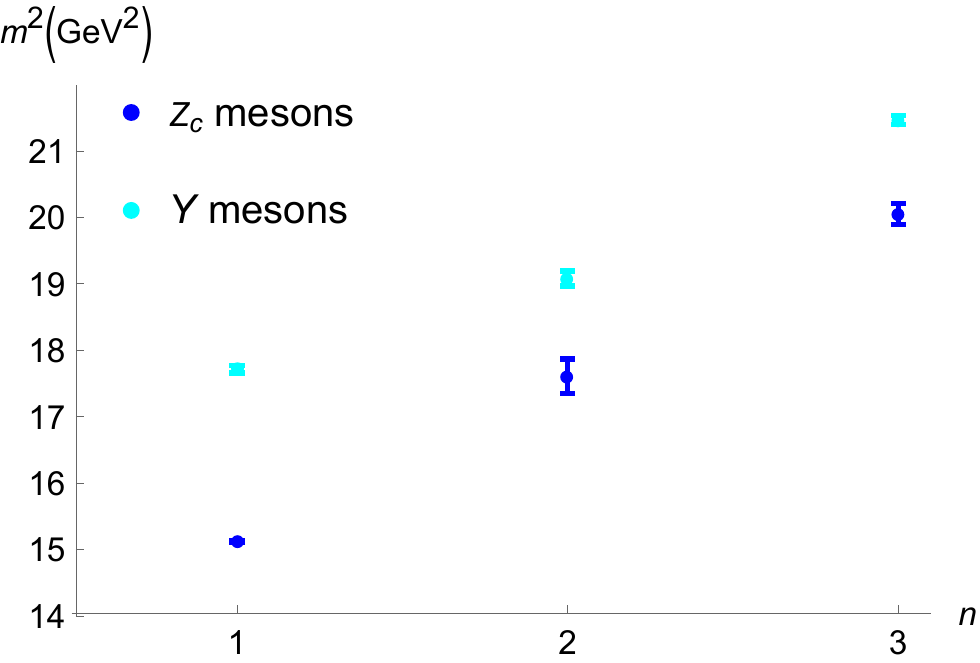}
	\caption{Experimental mass spectrum of the $Z_c$ and $Y$ exotic mesonic states as a function of radial excitations, for $n=1,\ldots,3$. The cyan points correspond to the $Z_c(1S)=Z_c(3900)$, $Z_c(2S)=Z_c(4200)$, and $Z_c(3S) = Z_c(4430)$ states, whereas the blue points depict the $Y(1S)=Y(4260)$, $Y(2S)=Y(4360)$, and $Y(3S)=Y(4660)$ states.}
	\label{tudonmn1}
\end{figure}

\begin{figure}[h]
	\centering
	\includegraphics[width=6.8cm]{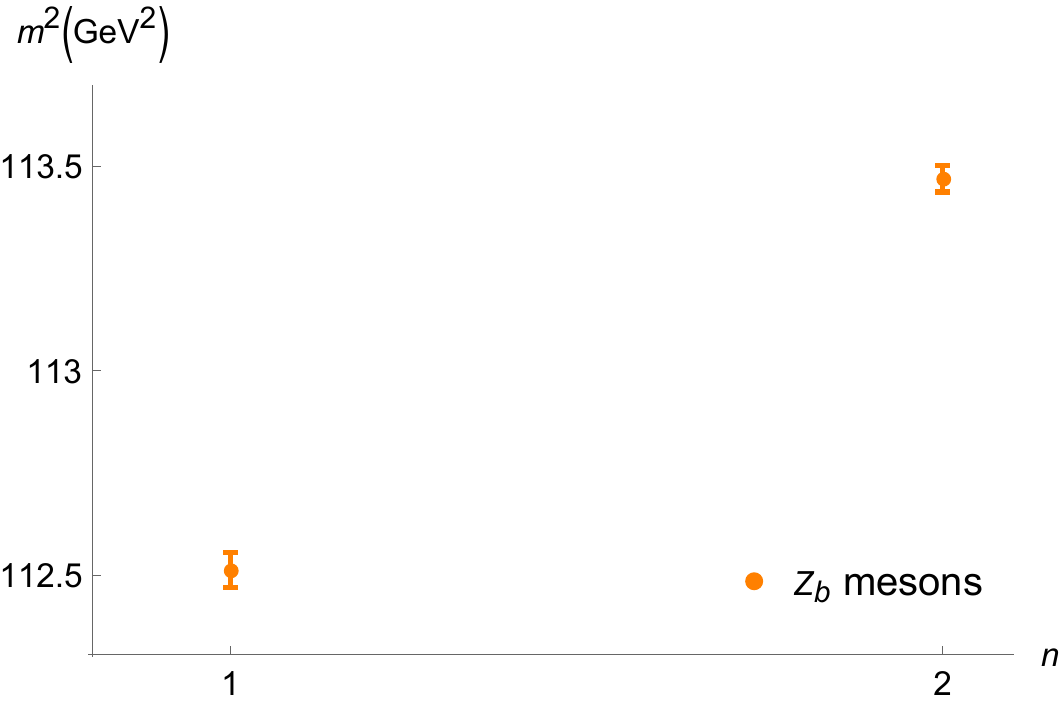}
	\caption{Experimental mass spectrum of the $Z_b$ exotic mesonic states as a function of radial excitations, for $n=1, 2$ (respectively corresponding to the $Z_b(1S)=Z_b(10610)$, and $Z_b(2S) =
 Z_b(10650)$ resonances.}
	\label{tudonmn2}
\end{figure}

\begin{table}[h]
\begin{center}
\begin{tabular}{||c|c||c|c||}
\hline\hline
$n$ & State & $M_{\scalebox{.65}{\textsc{Experimental}}}$ (MeV) & $M_{\scalebox{.65}{\textsc{Theory}}}$ (MeV) \\
    \hline\hline
\hline
1 &\;$Z_b(10610)$ & $10607.2\pm2.0$ & $10346.7$ \\ \hline
2 &\;$Z_b(10650)$ & $10652.2\pm1.5$ & $10696.6$ \\ \hline
\hline\hline
\end{tabular}
\caption{
Experimental and theoretical mass spectrum of $S$-wave resonances in the $Z_b$ meson family. For the masses in the fourth column, obtained by solving Eq. \eqref{schr}, $\Delta=7$, $\beta= 0.863$ and $\kappa = 11.649$ GeV.} \label{scalarmasses4}
\end{center}
\end{table}
Experimental data in Tables \ref{scalarmasses1} -- \ref{scalarmasses4} are shown in Figs. \ref{tudonmn} -- \ref{tudonmn2}.

\section{Mass spectroscopy of vector heavy-quark mesons from DCE}
\label{sec2}
The DCE can be computed starting with the energy density $\upepsilon(\vec{r})$, for $\vec{r}=(x^1,\ldots,x^d)\in\mathbb{R}^d$. 
Taking the spatial part $\vec{q}$ of the momentum $q$, the protocol to derive the DCE is based on the energy density in the momentum space, via a Fourier transform,
\beq\label{fou}
\upepsilon(\vec{q}) = \frac{1}{(2\pi)^{d/2}}\int_{\mathbb{R}^d}\,\upepsilon(\vec{r})e^{-i\vec{q}\cdot \vec{r}}\,{d}^d x.\eeq
Wave modes correspond to a probability distribution, $\mathbb{P}$, describing the power spectrum associated with the wave
mode, \cite{Gleiser:2018kbq} $
\mathbb{P}\left(\vec{q}\,\vert\, {\rm d}^dq\right)\propto \left|\upepsilon(\vec{q})\right|^{2}{d}^dq$. The modal fraction is then introduced as  
\cite{Gleiser:2012tu} 
\begin{eqnarray}
\upvarepsilon(\vec{q}) = \frac{\left|\upepsilon(\vec{q})\right|^{2}}{ \int_{\mathbb{R}^d} \left|\upepsilon(\vec{q})\right|^{2}{d}^d{q}}.\label{modalf}
\end{eqnarray}
The weight of information that is necessary to encode $\upvarepsilon$, with respect to wave modes, is calculated by the DCE,
\begin{eqnarray}
{\rm DCE}_{\upepsilon}= - \int_{\mathbb{R}^d}{\upvarepsilon^{\scalebox{.73}{$\,\triangleleft$}}}(\vec{q})\ln  {\upvarepsilon^{\scalebox{.73}{$\,\triangleleft$}}}(\vec{q})\,{d}^dq,
\label{confige}
\end{eqnarray}
where {$\upvarepsilon^{\scalebox{.73}{$\,\triangleleft$}}(\vec{q})=\upvarepsilon(\vec{q})/\upvarepsilon^{\scalebox{.68}{max}}(\vec{q})$}, and ${\upvarepsilon^{\scalebox{.68}{max}}(\vec{q})}$ denotes the maximal value of {$\upvarepsilon$}. 
To compute the DCE of heavy-quark exotic mesons, $d=1$ is chosen in Eqs. (\ref{fou}) -- (\ref{confige}), since the boundary where QCD lives has codimension one with respect to the AdS${}_5$ bulk. The energy density is the time component of the energy-momentum tensor
 \begin{equation}
{ \!\!\!\!\upepsilon\!=\! \frac{2}{\sqrt{ -|g| }}\!\! \left\{\!\frac{\partial\! \left(\sqrt{-|g|}{\mathfrak{L}}\right)}{\partial{g^{00}}} \!-\!\frac{\partial}{\partial{ x^\rho }} \!\!\left[ \frac{\partial\! \left(\sqrt{-|g|} {\mathfrak{L}}\right)}{\partial\!\left(\frac{{\scalebox{.79}{$\,\partial$}} g^{00}}{{\scalebox{.79}{$\,\partial$}}x^\rho}\right)}\right].
 \right\}}
 \label{em1}
 \end{equation}
Regarding the information underlying AdS/QCD, the experimental mass spectra of vector heavy-quark exotic mesons in PDG \cite{pdg} can serve as a substrate to extrapolate the mass spectra of higher excitations. This protocol is more reliable than the theoretically predicted mass spectra, obtained when the masses of meson states are seen as eigenvalues of Eq. (\ref{schr}). The experimental masses of vector heavy-quark exotic mesons are used together with the DCE (\ref{fou}) -- (\ref{confige}), evaluated as a function of the meson excitation numbers and their mass spectra. 
The DCE of exotic mesonic states in each of these families can be only calculated using numerical integration methods. 
The results are shown in Tables \ref{scalarmasses5} -- \ref{scalarmasses8}.\hspace*{-0.4cm}
\begin{table}[h]
\begin{center}
\begin{tabular}{||c|c||c||}
\hline\hline
$n$ & State & DCE (nat\footnote{Natural unit of information.}) \\
    \hline\hline
\hline
1 &\;$Z_c(3900)$ & 3.4811 \\ \hline
2 &\;$Z_c(4200)$ & 4.4227 \\ \hline
3& \;$Z_c(4430)$& 5.1359 \\ \hline
\hline\hline
\end{tabular}
\caption{DCE of $Z_c$ exotic meson resonances.} \label{scalarmasses5}
\end{center}
\end{table}
\hspace*{-0.4cm}
\begin{table}[h]
\begin{center}
\begin{tabular}{||c|c||c||}
\hline\hline
$n$ & State & DCE (nat) \\
    \hline\hline
\hline
1 &\;$Y(4260)$ & 5.2920 \\ \hline
2 &\;$Y(4360)$ & 7.7146 \\ \hline
3 &\;$Y(4660)$ & 11.0891 \\\hline
\hline\hline
\end{tabular}
\caption{DCE of $Y$ exotic meson resonances.} \label{scalarmasses6}
\end{center}
\end{table}
\hspace*{-0.4cm}
\begin{table}[h]
\begin{center}
\begin{tabular}{||c|c||c||}
\hline\hline
$n$ & State & DCE (nat) \\
    \hline\hline
\hline
1 &\;$\pi_1(1400)$ & 3.0415 \\ \hline
2 &\;$\pi_1(1600)$ & 3.6763 \\ \hline
3& \;$\pi_1(2015)$ & 4.2942  \\\hline
\hline\hline
\end{tabular}
\caption{DCE of $\pi_1$ exotic meson resonances.} \label{scalarmasses7}
\end{center}
\end{table}

\begin{figure}[h]
	\centering
	\includegraphics[width=8.4cm]{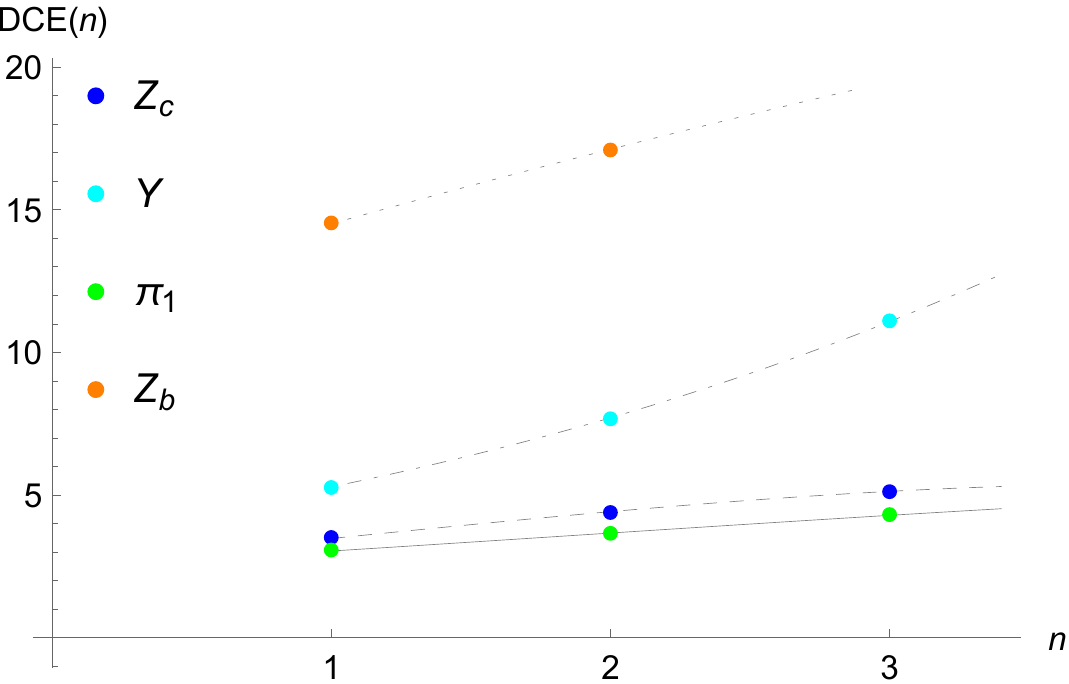}
	\caption{{DCE of exotic meson families as a function of the excitation levels.
The DRTs (\ref{itp1}) -- (\ref{itp4}) are plotted as: the dashed line, for the $Z_c$ exotic mesons (blue points); as the dot-dashed line, for the $Y$ exotic mesons (cyan points); as the dotted line, for the $Z_b$ exotic mesons (orange points); and as the solid line for the $\pi_1$ exotic mesons (green points).}}
	\label{cen1}
\end{figure}

\begin{table}[h]
\begin{center}
\begin{tabular}{||c|c||c||}
\hline\hline
$n$ & State &DCE (nat) \\
    \hline\hline
\hline
1 &\;$Z_b(10610)$ & 14.5121 \\ \hline
2 &\;$Z_b(10650)$ & 17.1230 \\ \hline
\hline\hline
\end{tabular}
\caption{DCE of $Z_b$ exotic meson resonances.} \label{scalarmasses8}
\end{center}
\end{table} DCE Regge-type trajectories (DRTs) are interpolation curves in the plot of the DCE as a function of the excited heavy-quark exotic meson states. With DRTs, the mass spectra of the next generation of $Z_c$, $Y$, $\pi_1$, and $Z_b$ exotic mesons can be extrapolated. 
 Fig. \ref{cen1} illustrates  interpolations of the data displayed in Tables \ref{scalarmasses5} -- \ref{scalarmasses8}, respectively for the $Z_c$, $Y$, $\pi_1$, and $Z_b$ exotic meson families. Each one of the DRTs are individually represented below:
\begin{subequations}
\begin{eqnarray}\label{itp1}
{\rm DCE}_{Z_c}(n)\!\!&\!=\!&\! -0.0365n^3\!+\!0.1043 n^2\!+\!0.8827 n\nonumber\\&&\qquad\qquad\qquad\qquad+2.5294
\\
{\rm DCE}_{\;Y}(n) \label{itp2}\!\!&\!=\!&\! 0.0318n^3\!+\!0.2841 n^2\!+\!1.3449 n\nonumber\\&&\qquad\qquad\qquad\qquad+3.6291,\\
{\rm DCE}_{\pi_1}(n) \label{itp3}\!\!&\!=\!&9.6456\times 10^{-3}n^3\!+\!0.0528 n^2\!+\!0.5389n\nonumber\\&&\qquad\qquad\qquad\qquad+2.4578,\\
{\rm DCE}_{Z_b}(n)\label{itp4} \!\!&\!=\!&\! -0.1207n^2\!+\!2.9721n+11.6586.
  \end{eqnarray}
  \end{subequations} Cubic interpolation for Eqs. (\ref{itp1}) - (\ref{itp3}), and quadratic interpolation for (\ref{itp4}) assure a supremum of $0.05\%$ root-mean-square deviation (RMSD).

\begin{figure}[h]
	\centering
	\includegraphics[width=7.8cm]{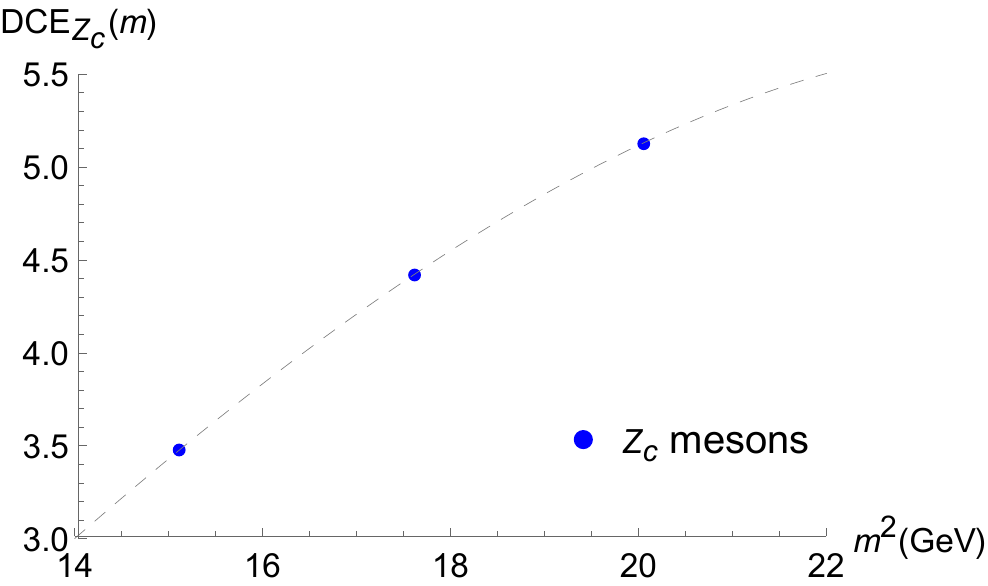}
	\caption{DCE of the $Z_c$ exotic meson family, for $n=1,\ldots,3$ (respectively corresponding to the $Z_c(1S)=Z_c(3900)$, $Z_c(2S)=Z_c(4200)$, and $Z_c(3S) = Z_c(4430)$ excitations in \cite{pdg}) as a function of the mass.
The massive DRT is displayed as the dashed line.}
	\label{cem11}
\end{figure}

\begin{figure}[h]
	\centering
	\includegraphics[width=7.8cm]{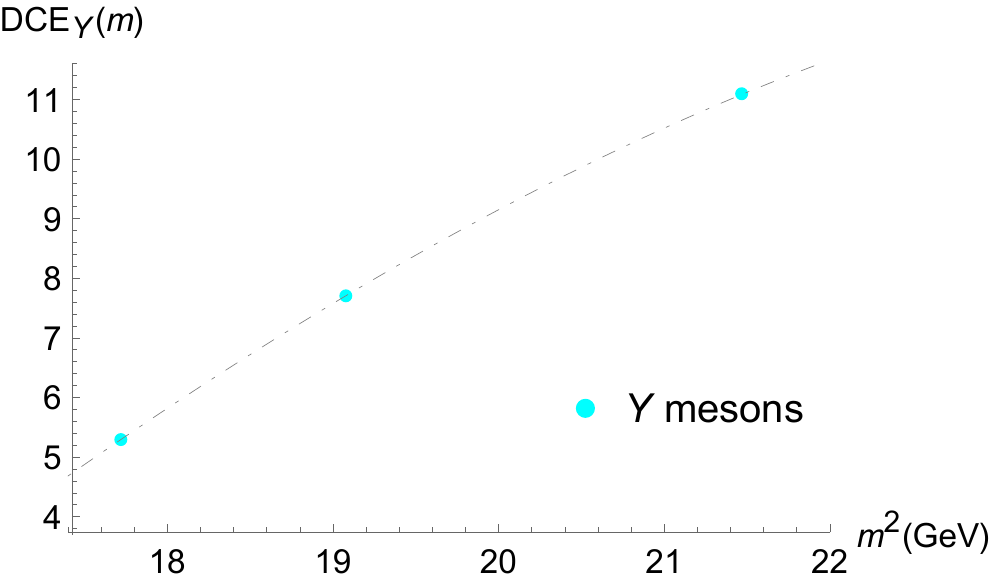}
	\caption{DCE of the $Y$ meson family, for $n=1,\ldots,3$ (respectively corresponding to the $Y(1S)=Y(4260)$, $Y(2S)=Y(4360)$, and $Y(3S)=Y(4660)$ excitations in \cite{pdg}) as a function of the mass.
The massive DRT is displayed as the dot-dashed line.}
	\label{cem12}
\end{figure}

\begin{figure}[h]
	\centering
	\includegraphics[width=7.8cm]{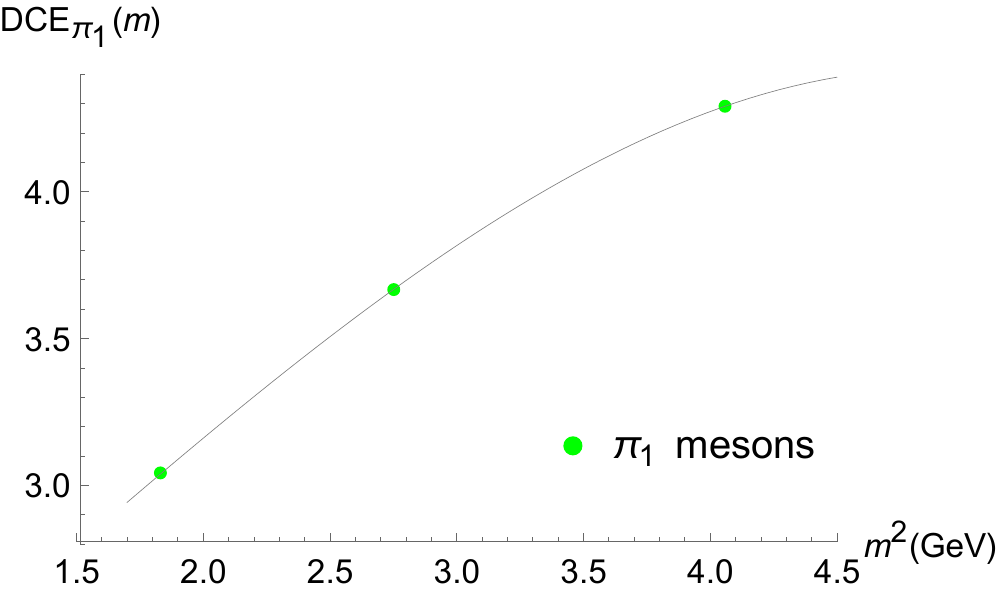}
	\caption{DCE of the $\pi_1$ exotic meson states, for $n=1,\ldots,3$ (respectively corresponding to the $\pi_1(1S) =\pi_1(1400)$, $\pi_1(2S) = \pi_1(1600)$ and $\pi_1(3S) = \pi_1(2015)$ excitations in \cite{pdg}) as a function of the mass.
The massive DRT is displayed as the gray line.}
	\label{cem13}
\end{figure}

\begin{figure}[h]
	\centering
	\includegraphics[width=8cm]{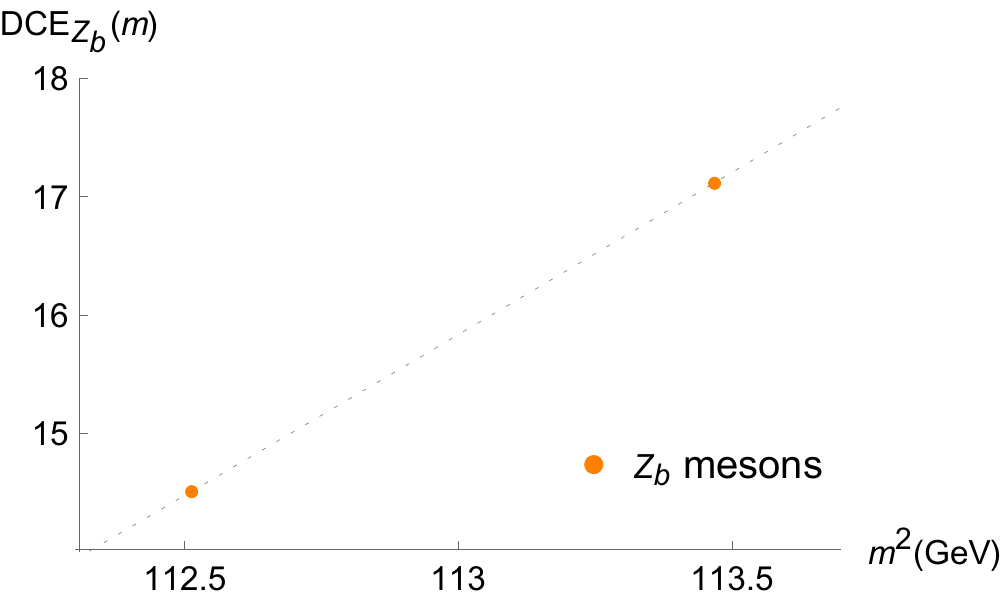}
	\caption{DCE of the $Z_b$ exotic meson states, for $n=1, 2$ (respectively corresponding to the $Z_b(1S)=Z_b(10610)$, and $Z_b(2S) =
 Z_b(10650)$ excitations in \cite{pdg}) as a function of the mass.
The massive DRT is displayed as the dotted line.}
	\label{cem14}
\end{figure}

The DCE that underlies heavy-quark exotica can be complementarily depicted as a function of the experimental mass spectra of the exotic meson states. Massive DRTs are then constructed as interpolation curves in the respective graphics of the $Z_c$, $Y$, $\pi_1$, and $Z_b$ mesonic $S$-wave exotic excitations \cite{pdg}.
The respective outcomes, for each exotic family, are pointed up in Figs. \ref{cem11} -- \ref{cem14}. The corresponding massive DRTs have explicit interpolation expressions given by Eqs. (\ref{itq11}) -- (\ref{itq14}). 
\noindent The massive DRTs, relating the DCE of the vector heavy-quark mesons to their squared mass spectra, $m^2$ (GeV${}^2$), respectively for the $Z_c$, $Y$, $\pi_1$, and $Z_b$ exotic meson states, are given by 
\begin{subequations}
\begin{eqnarray}
\label{itq11}
{\rm DCE}_{Z_c}(m)&=& -0.1150\, m^4+1.2849 m^2+2.3100,\\
 \label{itq12}
\!\! {\rm DCE}_{Y{\;}}(m) &=& 3.2346\times 10^{-3} m^6 +0.0916 m^4\nonumber\\&& \qquad\qquad
+1.6889 \,m^2 -35.4158,\\
 \label{itq13}
{{\rm DCE}_{\pi_1}(m)} &=& -0.01941 m^6 +0.0749 m^4+0.6497 \,m^2\nonumber\\&&\quad
 \qquad\qquad\qquad\qquad+1.7164,\\
 \label{itq14}
{{\rm DCE}_{Z_b}(m)} &=&  6.1911\times \!10^{-5} m^6 + 2.5389\times \!10^{-3} m^4 \nonumber\\&&\qquad\qquad
-0.2168\,m^2 -81.4167,
  \end{eqnarray}\end{subequations} In Eq. (\ref{itq11}) quadratic interpolation polynomial in the $m^2$ variable was used, whereas cubic ones were employed in Eqs. (\ref{itq12}) -- (\ref{itq14}), to drive the RMSD within a $0.06\%$ interval.

Eqs. (\ref{itp1}) -- (\ref{itp4}) yields the DCE of high excited exotic resonances in each family of heavy-quark exotica. Later, replacing the respective values of the DCE in Eqs. (\ref{itq11}) -- (\ref{itq14}), they can be algebraically solved to get the mass spectrum of heavy-quark exotic meson higher excitations. 
Starting with the $Z_c$ exotic meson family, the mass of the $Z_c(4S)$ state can be derived by replacing $n=4$ in (\ref{itp1}), yielding DCE$(Z_c(4S)) = 5.6099$ nat. It is then used on the left hand side of Eq. (\ref{itq11}), yielding  $m_{Z_c(4S)}=4.7850$ GeV. Performing the same routine for $n=5$ provides DCE($Z_c(5S)) = 5.8599$ nat. When Eq. (\ref{itq11}) is worked out for it the $Z_c(5S)$ exotic meson resonance has mass $m_{Z_c(5S)}= 4.9627$ GeV.
Table
\ref{scalarmasses10} summarizes these important results.
	\begin{table}[h!]
\begin{center}\begin{tabular}{||c|c||c|c||}
\hline\hline
$n$ & State & $M_{\scalebox{.65}{\textsc{Experimental}}}$ (MeV) & $M_{\scalebox{.65}{\textsc{Theory}}}$ (MeV) \\
    \hline\hline
\hline
1 &\;$Z_c(3900)$ & $3888.4\pm2.5$ & $3816.3$ \\ \hline
2 &\;$Z_c(4200)$ & $4196.0^{+31}_{-29}$ & $4213.9$ \\ \hline
3& \;$Z_c(4430)$ & $4478.0^{+15}_{-18}$ & $4551.4$ \\\hline
4${}^\diamond$& \;$Z_c(4S)\;$& ----- & 4785.0${}^\triangle$ \\\hline
5${}^\diamond$& \;$Z_c(5S)\;$& ----- & 4962.7${}^\triangle$  \\\hline
\hline\hline
\end{tabular}
\caption{Mass spectrum of high excited (indicated by a  diamond) $Z_c$ exotic meson resonances. The masses for the $n=4, 5$ states in the fourth column evidenced with `` ${}^\triangle$ '' result from the use of DRTs (\ref{itp1}, \ref{itq11}). } \label{scalarmasses10}
\end{center}
\end{table}

The DRTs (\ref{itp2}, \ref{itq12}) for $Y$ exotic meson states can be now utilized to get the mass of the $Y(4S)$ exotic state. Substituting $n=4$ in (\ref{itp2}) implies that the DCE$(Y(4S)) = 15.5909$ nat. Eq. (\ref{itq12}) can be therefore solved, whose solution yields $m_{Y(4S)}= 5.1027$ GeV. Accomplishing a similar routine to $n=5$, employing the DRT (\ref{itp2}) means that DCE$(Y(5S) =  21.4335$ nat. Consequently, solving Eq. (\ref{itq12}) after substituting this value on its left hand side implies that $m_{Y(5S)}= 5.2028$ GeV. 
The obtained results are summarized in Table
\ref{scalarmasses20}.

\begin{table}[h!]
\begin{center}
\begin{tabular}{||c|c||c|c||}
\hline\hline
$n$ & State & $M_{\scalebox{.65}{\textsc{Experimental}}}$ (MeV) & $M_{\scalebox{.65}{\textsc{Theory}}}$ (MeV) \\
    \hline\hline
\hline
1 &\;$Y(4260)$ & $4209.1\pm6.8$ & $4228.3$ \\ \hline
2 &\;$Y(4360)$ & $4368.0\pm13.0$ & $4577.3$ \\ \hline
3 & \;$Y(4660)$ & $4633.0\pm7.0$ & $4871.8$ \\\hline
4${}^\diamond$& \;$Y(4S)$& ----- & 5102.7${}^\triangle$ \\\hline
5${}^\diamond$& \;$Y(5S)$& ----- & 5202.8${}^\triangle$  \\\hline
\hline\hline
\end{tabular}
\caption{Mass spectrum of high excited (indicated by a  diamond) $Y$ exotic meson resonances. The masses for the $n=4, 5$ states in the fourth column evidenced with `` ${}^\triangle$ '' result from the use of DRTs (\ref{itp2}, \ref{itq12}).} \label{scalarmasses20}
\end{center}
\end{table}
\noindent

For deriving the mass spectrum of $\pi_1$ exotic resonances with a higher excitation level, let us start analyzing the $\pi_1(4S)$ exotic meson resonance. When $n=4$ is replaced in the DRT (\ref{itp3}), the DCE $\pi_1(4S) = 4.8421$ nat is obtained. If this value is substituted on the left-hand side of the DRT (\ref{itq13}), solving it yields $m_{\pi_1(4S)}= 2.2707$ GeV.
For $n=5$ in Eq. (\ref{itp3}) implies DCE$\pi_1(5S)=5.2685$ nat. Therefore replacing it on the left hand side of (\ref{itp3}) yields $m_{\pi_1(5S)}= 2328.1$ MeV.
These results are displayed in Table
\ref{scalarmasses30}. The $X(2210)$ element, with mass $2210^{+79}_{-21}$ MeV is a potential candidate for the $\pi_1(4S)$ state, whereas the $X(2340)$ state with mass $2340\pm 20$ MeV represents a candidate for the $\pi_1(5S)$ state \cite{pdg}.

\begin{table}[h!]
\begin{center}\begin{tabular}{||c|c||c|c||}
\hline\hline
$n$ & State & $M_{\scalebox{.65}{\textsc{Experimental}}}$ (MeV) & $M_{\scalebox{.65}{\textsc{Theory}}}$ (MeV) \\
    \hline\hline
\hline
1 &\;$\pi_1(1400)$ & $1354.0\pm25.0$ & $1351.7$ \\ \hline
2 &\;$\pi_1(1600)$ & $1660.0^{+15}_{-11}$ & $1646.6$ \\ \hline
3 &\;$\pi_1(2015)$ & $2014.0\pm20.0$ & $1901.7$ \\ \hline
4${}^\diamond$& \;$\pi_1(4S)$& ----- & 2270.7${}^\triangle$ \\\hline
5${}^\diamond$& \;$\pi_1(5S)$& ----- & 2328.1${}^\triangle$  \\\hline
\hline\hline
\end{tabular}
\caption{Mass spectrum of high excited (indicated by a  diamond) $\pi_1$ exotic meson resonances. The masses for the $n=4, 5$ states in the fourth column evidenced with `` ${}^\triangle$ '' result from the use of DRTs (\ref{itp3}, \ref{itq13}).} \label{scalarmasses30}
\end{center}
\end{table}
\noindent 

Now the $Z_b(3S)$ exotic meson resonance, can have its mass derived when inserting $n=3$ in Eq. (\ref{itp4}), yielding DCE$(Z_b(3S)) = 19.4886$ nat. Consequently, this value can be used on the left hand side of the DRT (\ref{itq14}), whose solution results $m_{Z_b(3S)}= 10.6928$ GeV. For the case $n=4$, Eq. (\ref{itp4}) implies that DCE($Z_b(4S)) = 21.6157$ nat. Finally the solution of Eq. (\ref{itq14}) when this value is substituted on its left hand side yields $m_{Z_b(4S)}= 10.7276$ GeV. These results are reported in Table
\ref{scalarmasses40}.

\begin{table}[h!]
\begin{center}
\begin{tabular}{||c|c||c|c||}
\hline\hline
$n$ & State & $M_{\scalebox{.65}{\textsc{Experimental}}}$ (MeV) & $M_{\scalebox{.65}{\textsc{Theory}}}$ (MeV) \\
    \hline\hline
\hline
1 &\;$Z_b(10610)$ & $10607.2\pm2.0$ & $10346.7$ \\ \hline
2 &\;$Z_b(10650)$ & $10652.2\pm1.5$ & $10696.6$ \\ \hline
3${}^\diamond$& \;$Z_b(3S)\;$& ------ & 10692.8${}^\triangle$ \\\hline
4${}^\diamond$& \;$Z_b(4S)\;$& ------ & 10727.6${}^\triangle$ \\\hline
\hline\hline
\end{tabular}
\caption{Mass spectrum of high excited (indicated by a  diamond) $Z_b$ exotic meson resonances. The masses for the $n=3, 4$ states in the fourth column evidenced with `` ${}^\triangle$ '' result from the use of DRTs (\ref{itp4}, \ref{itq14}).} \label{scalarmasses40}
\end{center}
\end{table}

\section{epilogue and outlook}\label{iv}

 The mass spectroscopy of the $Z_c$, the $Y$, the $\pi_1$, and the $Z_b$ heavy-quark exotic meson families was studied and discussed in the context of the DCE-based approach to AdS/QCD. 
The DCE technique is robust and uses DRTs as interpolation curves of the experimental mass spectrum of vector heavy-quark exotic mesonic states in PDG \cite{pdg}. As a result, the mass spectrum of the next generation of heavy-quark exotic meson resonances was derived, when using Eqs. (\ref{itp1}) -- (\ref{itp4}) and Eqs. (\ref{itq11}) -- (\ref{itq14}) together, respectively to the $Z_c$, the $Y$, the $\pi_1$, and the $Z_b$ heavy-quark exotic meson families.

The DCE also provides useful information regarding the configurational stability of heavy-quark exotic meson resonances. Tables \ref{scalarmasses5} -- \ref{scalarmasses8} show the DCE respectively calculated for the $Z_c$, the $Y$, the $\pi_1$, and the $Z_b$ heavy-quark exotic meson families. As the DCE attains higher values for higher excitation numbers in each exotic family, it is evidence that exotic mesonic states with higher excitation numbers present phenomenologically lower predominance. Hence, the DCE also supplies data about the relative abundance and dominance of $S$-wave exotic resonances of lower excitation levels, in the $Z_c$, $Y$, $\pi_1$, and $Z_b$ exotic meson families. At lower excitation levels, the information content of exotic resonances presents a higher compression into the QCD system wave modes, according to Shannon's interpretation of information entropy. The results provided by the DCE of exotic meson resonances shown in this work also comply with the lower production of exotic meson resonances with higher excitation levels, when compared to the lower excitation level exotic mesonic states. 
We can conclude that the more excited the exotic mesonic state, the higher its configurational instability is. Correspondingly, in each of the $Z_c$, $Y$, $\pi_1$, and $Z_b$ families, more massive exotic meson resonances present a higher configurational instability.
This feature supports the universal role played by DCE-based approaches to AdS/QCD.

To include the novel $S$-wave exotic excitations shown in Tables \ref{scalarmasses10} -- \ref{scalarmasses40} to the already existing database, besides considering the mass spectra of $Z_c$, the $Y$, the $\pi_1$, and the $Z_b$ heavy-quark exotic mesons, other features might be also probed, as the peaks of these resonances, their thresholds, decay constants, and widths. Although Ref. \cite{Braga:2015jca} provides a hint on how to obtain some of these quantities in the case of standard quarkonia, more specific developments must be emulated with new techniques that encompass the heavy-quark exotic mesonic states here studied. Also, besides the exotic families here studied, pentaquarks consisting of $c\bar{c}$ and light quarks, as the $P_c(4380)$ and $P_c(4450)$ exotica, may be described in AdS/QCD, once an appropriate scale dimension is either derived or chosen for them. Other hadronic states that include hexaquarks might have also their mass spectrum also extrapolated by DRTs, as well as double-heavy open-charm exotica and other tetraquark composites.

The extrapolation of the DRTs, to derive the mass spectra of new candidates for the next generation of exotic mesonic excitations in the $Z_c$ and $Y$ families, also fills the mass region 4.7 -- 5.2 GeV, which lacks standard mesonic states. It represents one more step to distinguish between exotic and standard mesonic states. Besides, some mesonic states, as the $X(4500)$ and $X(4700)$, are $D$-wave $cs\bar{c}\bar{s}$ tetraquark resonances \cite{Chen:2016oma}. Hence, the AdS/QCD approach to $S$-wave resonances must be extended to encompass $D$-wave resonances. Consequently, the DCE-based methods here employed can be extended to other multiquark exotic meson families. 
The advancing data on heavy-quark
meson spectroscopy, detected in several running experiments, can also provide available tools for a more comprehensive analysis of QCD exotica.

As a last perspective, dissociation of exotic meson states in finite-temperature plasmas, also with chemical potential dependence, can be approached using the techniques developed in Refs. \cite{Braga:2020myi,Braga:2020hhs}.  
\paragraph*{Acknowledgments:} GK thanks to The S\~ao Paulo Research Foundation -- FAPESP (grant No. 2018/19943-6), and RdR expresses gratitude to FAPESP (Grants No. 2017/18897-8 and No. 2021/01089-1) and the National Council for Scientific and Technological Development -- CNPq (Grants No. 303390/2019-0 and No. 406134/2018-9), for partial financial support.

\end{document}